# Predict genome-scale fluxes based solely on enzyme abundance by a novel Hyper-Cube Shrink Algorithm

Zhengwei Xie[1*†], Tianyu Zhang[2*], Qi Ouyang[3†]

[1]Department of Pharmacology, School of Basic Medical Sciences, Peking University, 38 Xueyuan Lu, Haidian District, Beijing, 100191, China
[2]School of Life Science, Peking University, 5 Yiheyuan Lu, Haidian District, Beijing, 100191, China
[3]Condensed Matter Physics, School of Physics, Peking University, Beijing 100871, China; Center for Quantitative Biology and Peking-Tsinghua Center for Life Sciences, Peking University, Beijing 100871, China

*Equally contribution
†To whom correspondence should be addressed. E-mail: xiezhengwei@bjmu.edu.cn, qi@pku.edu.cn

One of the long-expected goals of genome-scale metabolic modeling is to evaluate the influence of the perturbed enzymes to the flux distribution. Both ordinary differential equation (ODE) models and the constraint-based models, like Flux balance analysis (FBA), lack of the room of performing metabolic control analysis (MCA) for large-scale networks. In this study, we developed a Hyper-Cube Shrink Algorithm (HCSA) to incorporate the enzymatic properties to the FBA model by introducing a pseudo reaction constrained by enzymatic parameters. Our algorithm was able to handle not only prediction of knockout strains but also strains with quantitative adjustment of expression level or activity. We first demonstrate the concept by applying HCSA to a simplest three-node network. Then we validate its prediction by comparing with ODE and with a synthetic network in *Saccharomyces cerevisiae* producing voilacein and analogues. Finally we show its capability of predicting the flux distribution in genome-scale networks by applying it to the sporulation in yeast.

**Introduction**

One of the major part of current system biology in the –omics era is developing genome-scale models. Integrating various data sets to achieve overall and precise modeling of metabolism and thus predicting the consequences upon genetic perturbations or small molecules' inhibition is one of the major goals. Given the complexity of enzymatic reactions and high dimensionality of metabolic networks, such goal remains a challenging task using the ordinary differential equation approach. On the other hand, under the premise that biochemical networks are evolutionary optimal to maximize their growth rate or other end product production, Flux Balance Analysis (FBA) predicts testable flux distribution at steady state based on only stoichiometric numbers under the framework of linear programing[1,2]. This modeling overcome the scale difficulties and precisely predicted the essentiality of enzymes. The genome-scale models, including E. coli[1], yeast[2,3], rodent[4] and human[5] etc, have been built up and well curated[6]. In FBA, constraint that metabolic networks are at a steady state is imposed by stoichiometry in a biochemical network. For a network that includes $m$ metabolites and $n$ reactions, the sum of production and consumption fluxes of each internal metabolites, weighted by the corresponding stoichiometry coefficient, is zero: $S \cdot v = 0$, $S$ is the $m \times n$ stoichiometric matrix and $v$ is the fluxes.

Though FBA succeeded in predicting the essentiality of genes and the maximum evolution capacity under rare carbon source like glycerol[7], its limitation is tremendous. First, the optimality assumption engendered through evolutionary pressure could be invalid for engineered strains or other strains that haven't experienced such selections. Second, lack of the parameters of enzymes hindered us to perform MCA. Thus we were not able to predict the consequences of perturbed enzymes.

To incorporate the enzymatic information to the constraint-based model, the Boolean on/off constraints have been imposed to the network by cutting off the transcriptional level of enzymes[8]. As an improved implementation, iMAT[9,10] and MPA[11] have been developed to estimate the phenotypes using enzyme expression data as cues. Still, the expression of enzymes was used as discrete states (high level, median level and low level). The mixed integer linear programming (MILP) has been employed to find out the most likely consistency between expression and flux states (either $< \varepsilon$ or $> \varepsilon$). However, the threshold of the expression level is arbitrary and it is qualitative rather than quantitative. If the algorithm of incorporating the enzyme's abundance quantitatively exists, then one would be able to calculate the influence of stepwise change of each enzyme to all fluxes. Such information may be used to improve the yield of the end products or find out the drug targets for curing diseases, such as diabetes mellitus and cancers etc. The action matrix of enzyme deletions versus all fluxes would also be regressed to the corresponding phenotype vector and then the key reactions determined this phenotype would be revealed.

To solve this problem, we herein developed a Hyper-Cube Shrink Algorithm (HCSA) to find an interior point that is uniquely defined by the enzyme profile. Instead of using biomass as the objective function, we formulated a pseudo-reshaping flux whose maximum value defines the minimum volume of a hyper cube, which intersect original solution space at a single point. We succeeded in not only reducing the complexities but also retained some key features of Michaelis–Menten system.

First, we demonstrate the algorithm using a simplest three-node network. Second, we compared the its ability with ODE models in a six-reaction network. Third, we compared the prediction with experimental results in voilacein pathway. Finally, we applied it in genome-scale networks in the sporulation of budding yeast, *Saccharomyces cerevisiae*. We gave a reasonable overall estimation of amino acids synthesis in eighteen time points and found the potential rate-limiting enzymes in different time points.

## Results

### The Formula of Hyper-Cube Shrink Algorithm

We used a row vector to express this N-dimensional set of parameters, named Enzyme Distribution Vector (EDV), which is a key input for our linear model to include dynamical consideration and achieves a high correlation between proteins and flux. The N-dimensional flux distribution vector (FDV) is output of HCSA, playing the role as a straightforward readout of cellular functional state under a specific transcriptional state. Our goal is to heal the gap between transcriptional data and testable network phenotypes. Mathematically it is a mapping from EDV ($V_e$) to FDV ($V_f$). The problem is to find out the right solution from the space defined by $S \cdot v = 0$ plus extra constraints, like the reversibility or maximum uptake flux of nutrients etc. In FBA, the biomass flux was optimized based on the assumption that microbes were evolved to grow fast in a competitive environment. Herein, we constructed a pseudo flux $V$ as the objective function, which defines a hyper cube that shrinks while maximizing the matching of FDV and EDV along the increase of $V$. Besides the steady state equality constraint mentioned above, pairs of inequality constraints are set to FDVs, in the form:

$$l_i + C_i V \leq v_i \leq u_i - (1 - C_i)V \qquad (1)$$

$v_i$ ($C_i$) is the $i$th component of FDV (EDV), $u_i$ is the upper bound of reaction $i$. Here is HCSA in standard formula of LP problem:

$$maximize\ V\ s.t.$$

$$[S_{m \times n} \quad 0_{m \times 1}] \begin{bmatrix} v_{n \times 1} \\ V_{1 \times 1} \end{bmatrix} = 0_{(n+1) \times 1} \qquad (2)$$

$$\begin{bmatrix} -I_{n \times n} & C_{n \times 1} \\ I_{n \times n} & 1_{n \times 1} - C_{n \times 1} \end{bmatrix} \begin{bmatrix} v_{n \times 1} \\ V_{1 \times 1} \end{bmatrix} \leq \begin{bmatrix} 0_{n \times 1} \\ 1_{n \times 1} \end{bmatrix} \qquad (3)$$

$$v_i \in [L_i\ U_i]$$

Started with original FBA formula (the solution space is the blue triangle in Figure 1A), we added two pseudo metabolites implemented by equation (3) to each reaction to restraint the dynamic lower and upper bound. All the pseudo metabolites were coordinated by a pseudo reaction $V$, where the coefficients ($C_i$) of pseudo metabolites in $V$ are the normalized abundance of each enzyme. When $V$ varies from 0 to $max\ V$, it defines hyper cube shrinkage (as shown in Figure 1F-G). The vertex of the cube closest to the origin moves at the speed defines by $C_i$ along each axis. The solution space for the entire problem is thus the cross-section between the hyper cube and polyhedron defined by original FBA.

We use the three-reaction toy model (Figure 1A) to show the concept. The equality and inequality constraints define a feasible space $\Phi$ in N-dimensional vector space collaboratively. $\Phi$ is an intersection of a polyhedron (Figure 1B) and a hyper cube (Figure 1C) whose volume is shrinking as the problem solver finds a higher and higher value of the formal parameter $V$ (yellow triangle in Figure 1F). Although not proven, $\Phi$ shrinks to a single point when V takes its maximum value (Figure 1G), since the edges of hyper cube are parallel to coordinate axis and the polyhedron is an oblique one which passes through the origin. It is noteworthy to remark that FDV is a balanced solution for the steady state of metabolic network and our EDV is outside $\Phi$ mathematically.

We show a solution $FDV = [1\ 5/9\ 4/9]$ of the toy model for $EDV = [8\ 5\ 4]$ as shown in Figure 1H. We obtained a FDV well correlated with EDV as shown in Figure 1I. The visualization of HSCA was shown in Supplementary Movie 1. We also varied EDV(2) from 0.5 to 5, followed by varying EDV(3) from 4 to 0.4. The phase transition curve of this change was shown in supplementary Figure 1 and supplementary Movie 2.

**A similar enzymatic dose response behavior compared with ODE model**

With the $K_{cat}$ implied in the model, we were able to simulate the network for different enzyme expression levels. To test the performance of HCSA, we used ordinary differential equation to model the six-reaction network, characterized by Michaelis constants ($K_m$'s) and catalytic rate constant ($k_{cat}$'s) of the enzymes. We varied the enzyme amount to the same extent in both models to compare the FDVs.

One of the main functions of our algorithm is its ability of dealing with the variation of enzyme expression levels. And considering the amount of enzyme will significantly influence the dynamic interaction between substrates and enzymes, we built a concise ODE model to test HCSA. The ODE system consists of six biological reactions, and all of them are supposed to be Michaelis–Menten ones, irreversible. Each enzyme was given a unique $k_{cat} - K_m$ combination (See supplementary Table I). Each time we altered one of enzyme amounts in ODE model to simulate a regulation of gene expression, and modify the corresponding component in EDV proportionally. The amounts of enzymes that catalyze number 2, 4, 6 reactions varied by fold from 0.1 to 10 (Figure 2A-C). The norm of six simulated FDVs are normalized to one so as to show the relative magnitudes of six flux are predicted in the same pattern in these two kinds of models, the shape of curves generated by ODE and HCSA were perfectly matched (Figure 2D-I). Overall, 77% of curves have correlation coefficients over 0.99(Figure 2J-L).

There is a scenario can be easily imaged. When the downstream enzymes were up-regulated to very high level, the flux of this branch will asymptotically approach to a limit number because of the bottleneck effect of upstream enzymes. Indeed, we observed this phenomenon under both frameworks (Figure 2D-I). And the excess of an enzyme leads to "substrate saturation". If we calculate the $\Delta v/\Delta E$ for each change pair, this ratio is always less than unit. It means the network is always resistant to enzyme perturbations, implying the system is robust. Our HCSA has captured this feature in essence.

Real biological dynamic networks might contain reversible reactions or allosteric feedback terms thus have a modified $K_m$, but our simulation showed that the strong positive correlation still exists, as shown in supplementary Figure 2. Much of our algorithm's prediction ability is attributed to its good consistency with classic ODE model, which gives it foundation to work with empirical subjects effectively.

**Prediction of flux distribution in violacein biosynthesis pathway in** *Saccharomyces*

*cerevisiae*

We applied our model to the violacein pathway in *Saccharomyces cerevisiae* to test its efficacy. As described previously, this branch of metabolic network includes nine metabolites, six enzyme catalyzed steps and two non-enzymatic reactions[12,13], as shown in Figure 3A. A set of promoters (pTDH3, pTEF1, pRPL18B, pRNR2 and pREV1) is used to regulate the expression level of the five key enzymes - VioA, VioB, VioC, VioD and VioE. The five promoters were able to produce a wide range of expression levels change. 191 samples of the expression library were used and their promoter genotypes are identified using TRAC[12]. HPLC analysis was employed to measure the pathway products, from which the flux distribution could be uniquely determined.

Such a promoter combination provides both a challenge for a computational model and a good opportunity to reveal its power. We use 7% (14 of 191) data groups as training sets and the rest for testing. Each time a promoter combination is chosen, and the strength of the promoters directs us to modify the components of the datum quantitatively, representing the fold change of gene expression level. And HCSA was employed to calculate ratios of the eight flux' magnitude were obtained, see supplementary text for details.

The absolute amplitude of the flux of the entire branch is determined by the enzyme expression of the first three reactions and their cross product. We used a multiple regression model to predict it, as shown in supplementary text and Figure 2. The parameters of this model were obtained through the training set. The ratios from HCSA multiply the amplitude gave us the final flux. We obtained very good consistency between HCSA simulation and measured flux. The result is visualized in Figure 3B – 3E, and the square of correlation coefficient is 0.76, 0.72, 0.51 and 0.84 for violacein, deoxyviolacein, proviolacein and prodeoxyviolacein respectively, as shown in Figure 3.

We were also curious about if we are able to validate the saturation curves in Figure 2D - 2I. Some single-variable subsets of experimental data were screened out. And here we choose one in which only enzyme VioD varies, whose expression level was modified throughout a range of almost one thousand folds This wide range of regulation results in a dramatic increase in production of violacein and proviolacein as well as a drop in the other two primary products. It looks like a seesaw was pulled up at one end causing pushed down at the other end. This redistribution of flux follows a "Seesaw" like mechanism, as shown in supplementary Figure 4. We noticed the saturation behavior mentioned before appeared again in both the experimental data and our simulation flux data, which means HCSA captured the key characteristics.

**Genome-scale network implementation of HCSA**

When the network was scaled up to genome-scale, we wanted to test if HCSA is still applicable. Here we chose Yeast iND750[3] as the model system. We used the ribosomal profiling data performed by Brar et. al[14] during the time course of meiosis, which occurs when diploid yeast cells were incubated in poor media and low temperature. The ribosomal profiling data covered 21 different time points at different stages of sporulation.

*HCSA gets more reasonable flux distribution compared to conversional methods*

We first compared the results obtained by FBA and HCSA in normal growth condition. In this analysis, we kept the biomass flux and restricted it in a reasonable range (0.3~maxflux). Then we maximized the pseudo flux V. We compared the results with previous measured flux by Moxley et. al[15] in yeast. Among 77 reactions with experimentally measured fluxes, 51 non-zero-fluxes have been obtained by FBA and 64 non-zeros have been obtained by HCSA. For instance, in FBA threonine synthesis has been catalyzed through glycine + acetaldehyde → L-threonine by Gly1, of which the specificity is too low to occur experimentally. In HCSA, the normal threonine synthetic pathway has been chosen according to the expression level. Other different reaction sets were listed in Supplementary Table IV. Using experimental fluxes as the reference, the correlation coefficient with HCSA fluxes and FBA fluxes are very similar (0.81 versus 0.81), while HCSA flux has a smaller *p*-value because it has more non-zero data points ($1e^{-15}$ versus $3e^{-13}$).

*HCSA predicts correct yields of end product*

In HCSA, the fluxes were determined by enzyme abundances. The biomass flux was thus not necessary. We split the biomass flux into 43 outflow reactions and used 0.5 as the pseudo expression level for all of them. After maximizing V, we obtained their outgoing rate, as shown in Figure 4A. The yield of biomass precursors correlated very well ($r^2 = 0.81$ in log scale, as shown in Figure 4A) with their corresponding biomass coefficient that determined from dry mass of yeast.

*Simulate flux from ribosomal profiles during sporulation*

FBA does not need the abundance data and the kinetic parameters. It was an advantage when all these data are missing. But now it's a shortcoming for taking advantage of genome-scale data. Covert et al developed an algorithm to cut the abundance data with a threshold then applied a Boolean constraint in FBA model[16]. In sporulation, a cell transformed to four spores but and did not grow in volume thus one could not optimize the biomass to obtain the EDV. We used the ribosomal profiling data measured by Brar et. al[14]. We first changed the medium conditions, including shutting off glucose uptake flux, turning on acetate uptake reaction and opening amino acids uptake reaction to the amount less than biomass coefficient. We split the biomass flux to subgroups, including a.a. (amino acids) synthetic reactions (consist of 20 amino acids using the coefficients same as in biomass), nucleotide synthetic reactions (consist of 4 nucleotides) and others (individual output flux or single metabolite). The EDV for a.a. flux was set to NULL so that this flux will be determined by relative enzymes rather than uptaken directly from the medium. We embedded the EDVs in various sporulation stages to virtual V and obtained corresponding FDVs.

The uneven expression profiles (Figure 4B) was converted to even flux profiles (Figure 4C) by HCSA, indicating the balance of the metabolites had de-noise effect. The median curve of EDVs showed four up-regulation peaks, which means the cells surfers continuous amino acids starvation.

The combined amino acid reaction showed a continuous trend during the entire sporulation (Figure 4D, red curve). The average expression level increased by ~4 folds, yield 2 folds increase of synthetic flux at time point E. The a.a. flux then decreased gradually until the lowest point K. At this point, cells were at the very beginning of metaphase I. After that a.a. flux increased gradually through metaphase I, anaphase I, metaphase II, anaphase II until spore formation.

By comparing the median EDV curve and a.a. flux curve in Figure 4D, we observed that when the median EDV just started to be up-regulated, the a.a. flux didn't response immediately (blue drop lines). In the latter part of the two curves, the efficacy $V_{aa}/E$ increased, showing the coordination between different pathways yield more efficient overall flux.

*Hom6 might be the rate-limiting enzyme*

MCA exposed the casual relationship between enzymes' abundance and fluxes. Because the abundance of enzyme were embedded in the model, we were able to calculate the metabolic control coefficient, which is defined as $\gamma_i = \Delta v/\Delta E$. The full metabolic control coefficient matrix $\Upsilon$ was visualized in Figure 5A. We found two genes catalyzing three reactions have positive $\gamma_i$, which means increasing $E_i$ cause increased a.a. flux. Taking time point B as the starting point, increase Hom6 and Gly1 by 1.2 fold was able to increase a.a. flux by 1.1 and 1.03 folds respectively, as shown in Figure 5A, column 2 row 1~3. Across all the conditions, Hom6 (catalyzing HSDyi & HSDxi) and Gly1 (catalyzing THRA) possess most positive $\gamma$. The control point in this figure means the enzyme is the rate-limiting step. One may expect that they all have low expression level. The expression level of these enzymes were shown in Figure 5B, we found most of these enzymes have been relatively high expressed, which means they are under tight regulation.

From Figure 4D, we found the total enzyme expression decreased but the a.a. flux increased. In Figure 5B, DDPA, DDPAm, PPNDH, HISTP and PHETA1 decreased in the latter part. One possible explanation for this contradiction is that cells uptake more amino acids from the SPO media. As shown in Figure 5C, the uptake flux of PRO, PHE, MET, LYS, LEU and ILE increased significantly in the latter part.

We also calculated the correlation between enzyme abundance data and the a.a. flux. The results were listed in supplementary Table VI. Hom6 has the best correlation ($r^2 = 0.8$) with a.a. flux among all the enzymes, the scatter plot is shown in supplementary Figure 5. It is noteworthy that not all the enzymes with good correlation have casual effect. The good correlation may come up with co-regulation. As shown in Figure 5A, we observed Hom6 has positive $\gamma$ for eight times, meaning the result does not depend on the conditions.

### Discussion

In this study, we developed an algorithm to incorporate enzyme information to the stoichiometric model, we call it hyper cube shrink algorithm (HCSA). Incorporation of enzyme activity is urgent for metabolic engineering and understanding / curing the metabolic disorder of diseases. In some genetically engineered bacterium strains, the expression of a specific enzyme can be quantitatively changed, which means we should adjust EDV quantitatively. This can not be done by FBA[1,17] or MOMA[18] because they can not handle the case of expression level modification. But this is what HCSA is for, interestingly, prediction of HCSA corresponds well with ODE models and experimental data in dynamic patterns.

Previously, Integrative Metabolic Analysis Tool (iMAT)[10] and IOMA[19] have been developed to incorporate expression data to the models. iMAT uses mixed integer linear programming (MILP), which requires extensive parallel computing resources. It maximizes the number of reactions whose flux is consistent with their expression level. For both expression data and predicted flux, the values have been simplified to Boolean numbers. In contrast, we use the exact values as the estimation cues. The overall flux of a branch was determined by the combined effect of all the enzymes. Thus, we obtained more reasonable flux distributions. IOMA is formulated as a quadratic programming problem requiring proteomic and metabolomics data. It required the turnover rate of enzymes and thus limited its application.

Another alternative method, CEF or mCEF[20,21], has been developed based on the elementary modes (EM) of the network to solve the genetic modification problem. It uses quadratic programming to find out the linear combination of EMs to resemble the gene expression profile. It has the followed drawbacks. First, EM is hard to be computed for larger networks and requiring manually merging the reactions to reduce the computation burden. Second, its solution is not unique, thus it's hard to compare the flux change upon enzyme changes. Third, it's not very intuitively clear mathematically.

Essentially, HCSA find a solution well correlated with fluxes, which is supported by experimental results. This correlation is not complete because of regulated enzyme kinetics. Thus, the expression data were just treated as cues for determining the metabolic flux of corresponding reactions. As the enzyme activity was treated as parameters in the model, the perturbed network can be easily simulated. Thus the metabolic control matrix was obtained. Further, relying on enzyme expression data to infer the metabolic flux eliminates the need of objective function. Rather, the effluxes can be predicted based on the EDV. Though not shown here, HSCA were easily be applied to large scale networks, like human network, which is comprised of more than 2000 reactions.

There are still some inconsistency between our prediction value and the empirical curve in Figure 3, which indicates the existence of some secondary interactions between enzymes and metabolites. The inconsistency between our prediction and the experimental data also comes from the intrinsic property of the promoters[22,23]. The scatter plot of RNA level versus protein level bears similar shape[24]. This phenomenon calls for an improvement of HCSA if one asks for more precise predictions. At the same time, it also interprets HCSA's power to tolerate inclusion of relatively trivial data when people are working on pathways with only incomplete knowledge.

**Figure 1. (A) – (F)** Take a one-node metabolite network as an extremely simplified example to illustrate the optimization principle underlying HCSA. **(B)** The solution space constrained by $v_1 = v_2 + v_3$ is within a plane (colored blue). **(C)- (F)** The direction of EDV is represented by a red arrow, showing how the lower bounds of fluxes increase. The direction of the black arrow shows how upper bounds decrease. Geometrically, the magnitudes of both arrows are proportional to the magnitude of pseudo reaction V and the orange cube is uniquely determined by them. **(C)** The state in which the value of objective function is still far below maximum. **(D)** The feasible solution space is defined as the intersection of the plane and cube (yellow triangle). **(E)** When objective function reaches its optimum, the length of two arrows reach maximum and the cube becomes smallest. So all the candidate points in it uniformly have a strong correlation to the EDV vector. **(F)** The optimal solution, FDV, is the single point still in the intersection of flux balance plane and the hyper cube. **(G)** Analysis of the optimum state in terms of the three coordinates respectively. The light green segments rebuild the spatial cube and the components of FDV lie either on the upper bounds or the lower bounds. **(H)** A strong correlation between the enzyme parameters and flux distributions is indeed established by HCSA.

**Figure 2**. The comparison of HCSA and ODE model on a branch-like metabolic network. **(A) – (C)** The topology of the network, thick arrow indicates the flux whose enzyme parameter

uniquely varies. **(D)- (F)** The dose response of fluxes upon enzymes changes from $10^{-1}$ to $10^{1}$ for reaction 2, 4, 6, simulated by HSCA. When the enzymes are at low concentration, the fluxes are linearly correlated with enzymes. When the enzyme is over dosed, the flux approaches to an asymptotic value. **(G)- (I)** The curves simulated by ODE model. **(J)- (L)** The comparison of ODE and HCSA fluxes predictions in which over 77% of results have correlation coefficients over 0.99.

**Figure 3.** Application of HCSA to quantitatively predict flux distribution in violacein pathway. (A) The violacein biosynthetic pathway in *Saccharomyces cerevisiae.* It's a metabolic network consists of 6 enzymatic and 2 non-enzymatic reactions. L-Trp, L-Tryptophan; PDVA, protodeoxyviolaceinic acid; PVA, protoviolaceinic acid; PDV, prodeoxyviolacein; DV, deoxyviolacein; PV, proviolacein; V, violacein. Bold arrows show the modified enzymes. (B-E) Model predictions for a test set of 191 unique promoter combinations. The overall Pearson correlation coefficient is 0.67 and it is 0.76 for violacein (red), 0.72 for deoxyviolacein (blue), 0.51 for proviolacein (green) and 0.84 for prodeoxyviolacein (cyan). **(F)** The expression level of VioD is uniquely modified throughout a wide range of one thousand folds experimentally, measured to be 0.0013, 0.0174 0.0918,1. **(G)** HCSA simulation behaves similarly.

**Figure 4.** Application of HCSA in the genome-scale metabolic network of *Saccharomyces cerevisiae* (A) The yield of biomass components obtained by HCSA correlated well with the corresponding coefficient in previous constraint-based model ($r = 0.85, p < 4e - 13$). Red dashed line is the fitted line. The expression curves (B) and the flux curves (C) across different experiments for all the enzymes in amino acids synthetic pathways. Green and red lines are the median curves. (D) The median curve of enzyme expression (green) and the coupled amino acids synthetic reaction (red) were merged for comparison. The peak of gene expression, a.a. flux and the lowest point of a.a. flux was highlighted by green, red and blue drop line respectively.

**Figure 5.** The sensitivity analysis of the amino acid synthetic network. (A) Metabolic control coefficient $\gamma_i = (v_{E=1.1*E_0} - v_{E=E_0})/v_{E=1.1*E_0}$ at different time points were shown in heat map. At each time point (columns, defined by Brar et. al.), at least one reaction was rate-limiting. The reactions (rows), like HSDyi/HSDxi/THRA, were rate-limiting in multiple conditions (columns). (B) The expression of the enzymes in all the time point was visualized. (C) The uptake rate of amino acids at different time point was shown in heat map. The down-regulated enzyme expression does not necessarily to be rate-limitting. For instance, the down-regulation of phenylalanine synthesis and histidine (red arrows in B) were compensated by increasing uptake rate in the late stage of sporulation (green arrows in C).

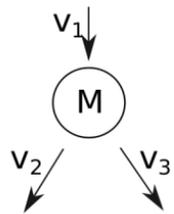
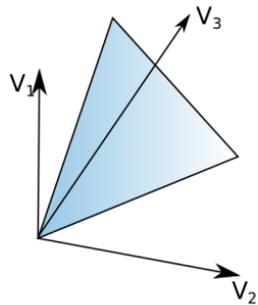
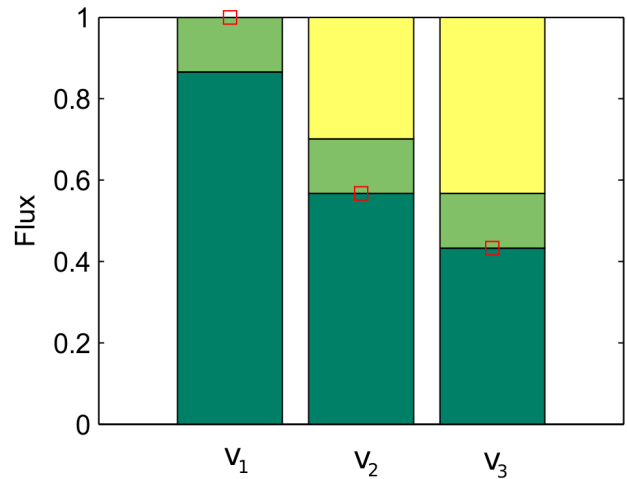
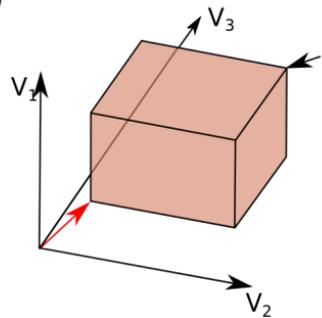
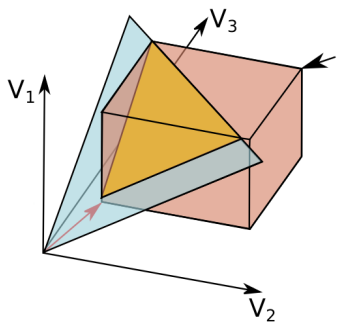
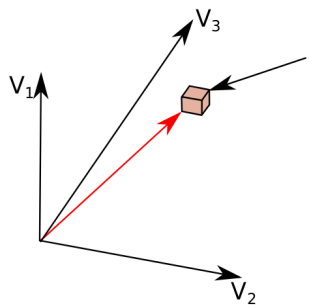
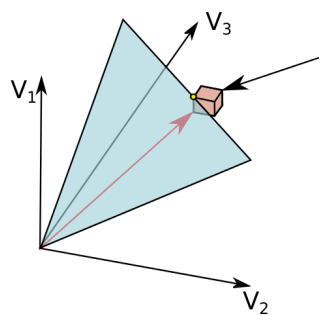
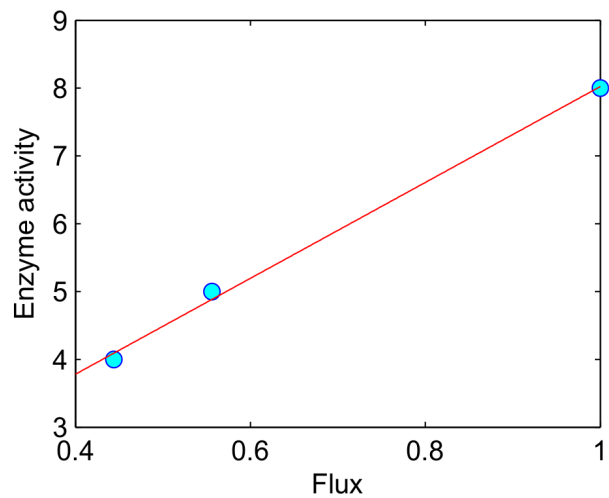

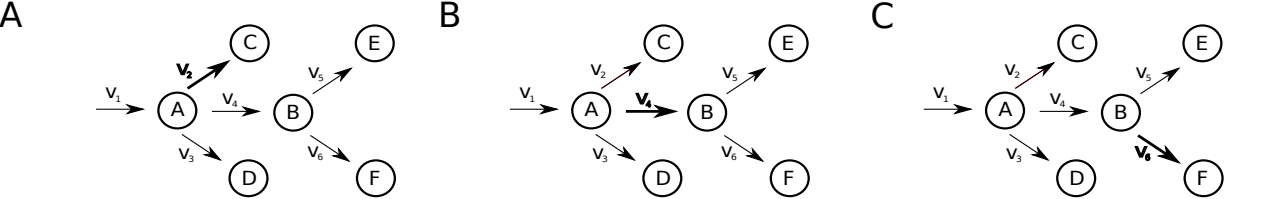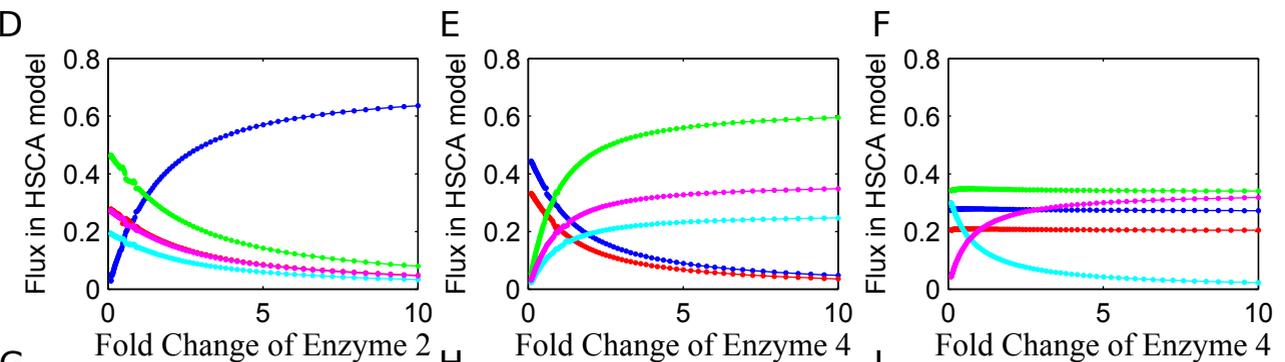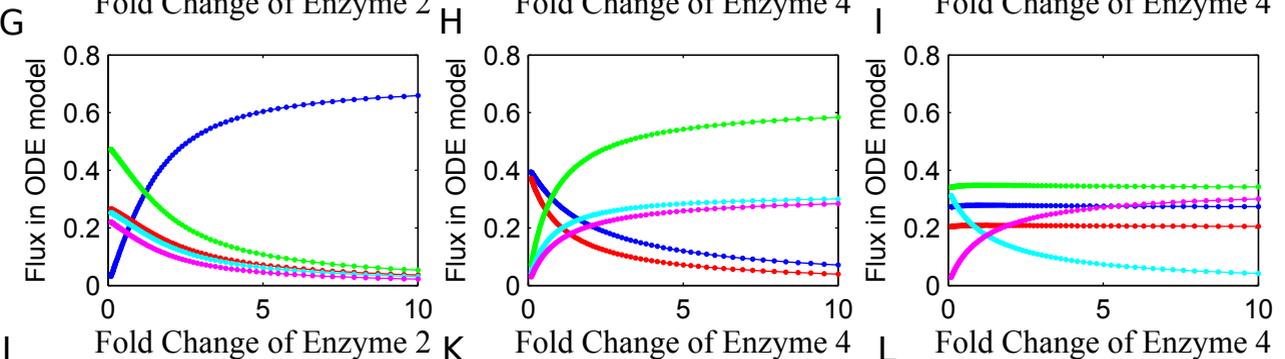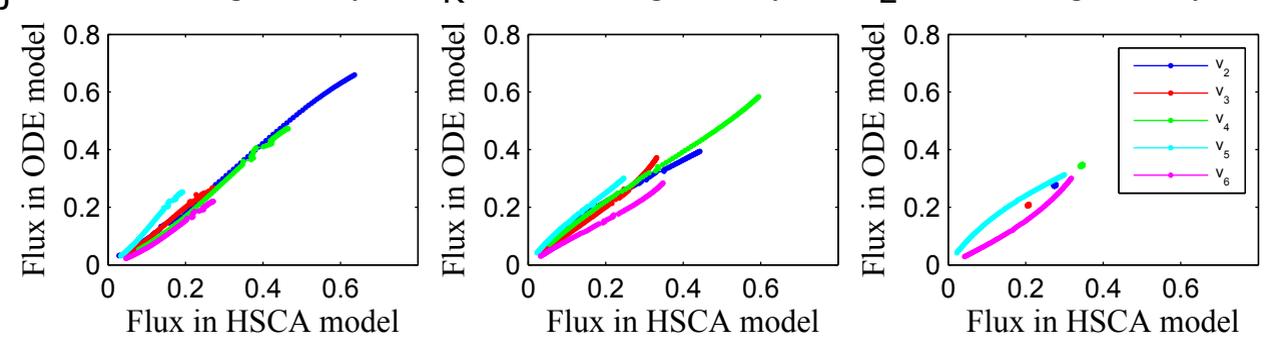

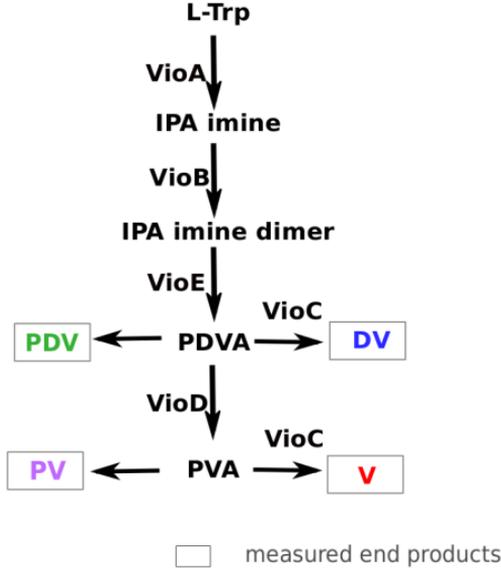

(A)

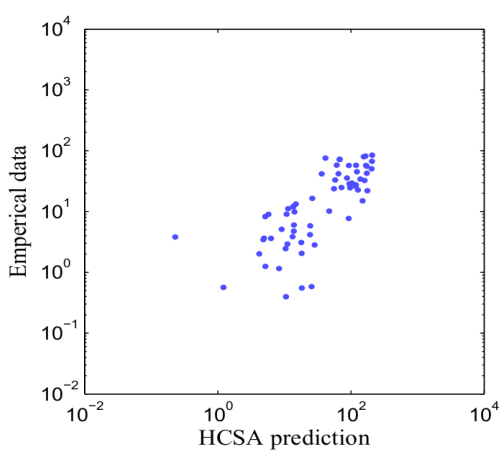

(B) deoxyviolacein

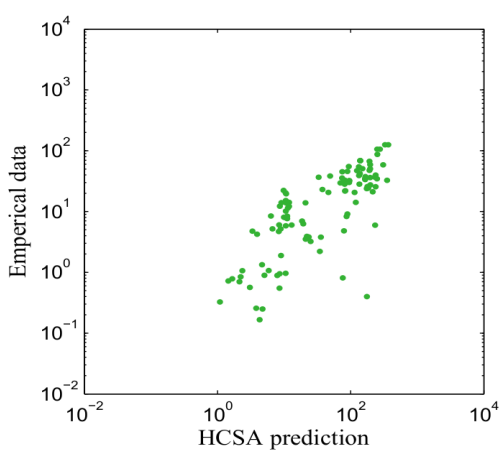

(C) prodeoxyviolacein

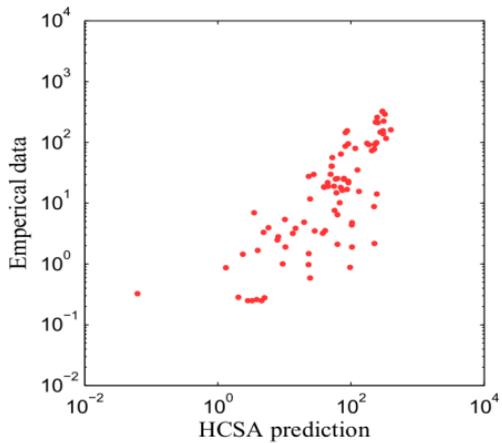

(D) violacein

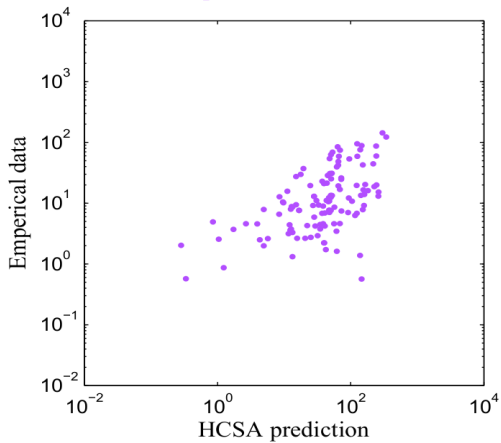

(E) proviolacein

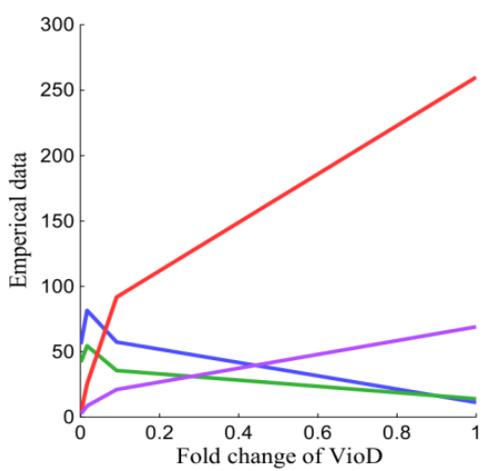

(F)

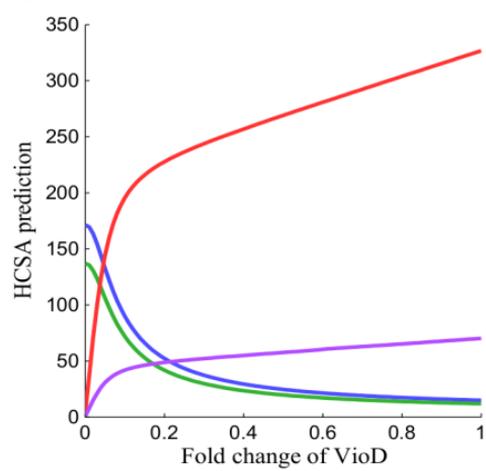

(G)

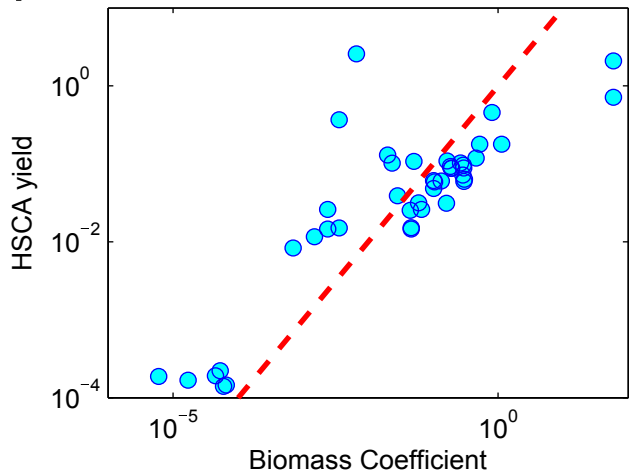
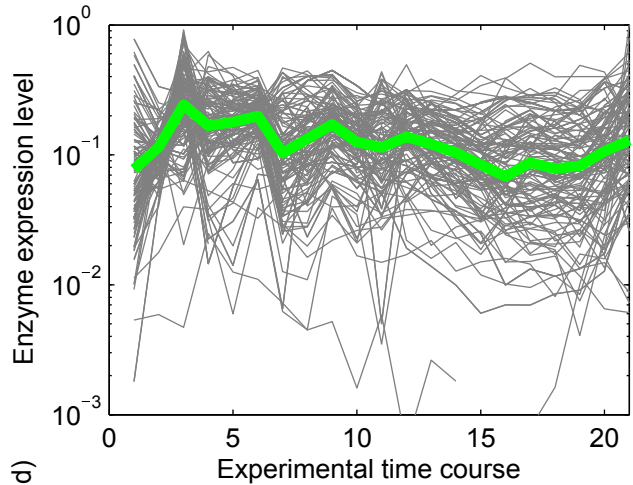
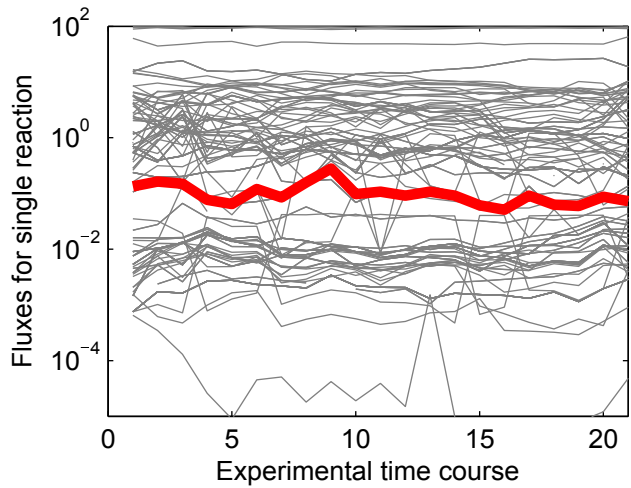
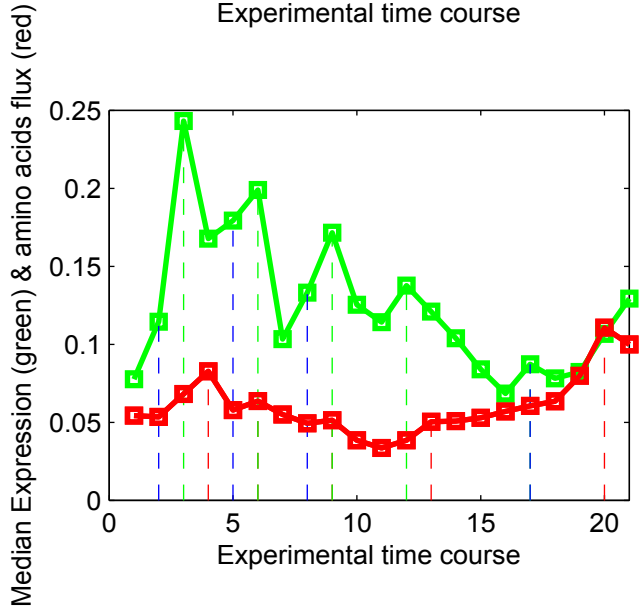

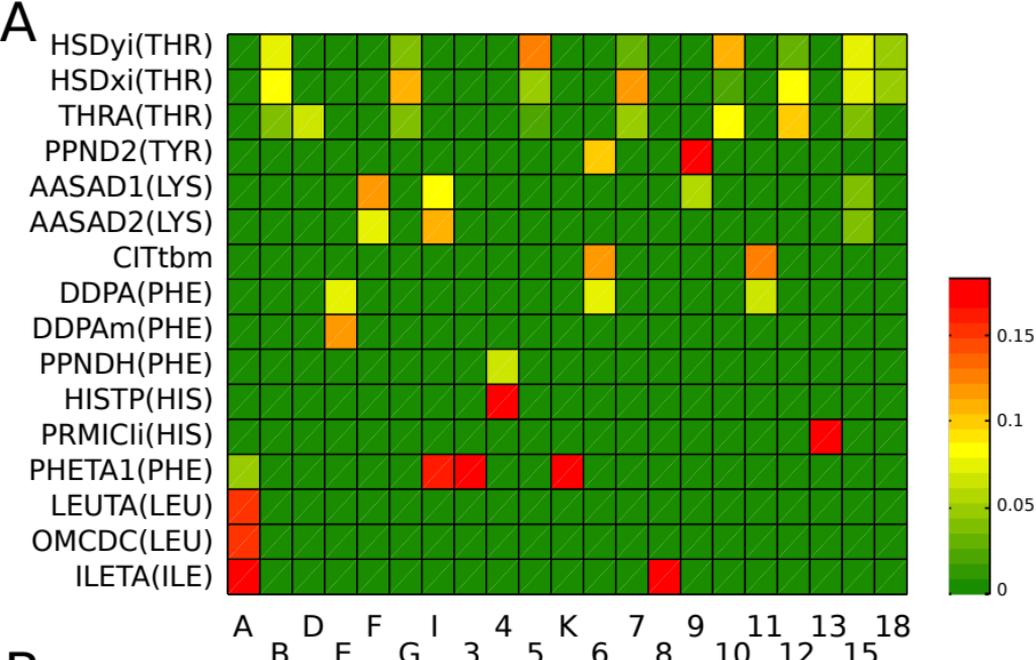
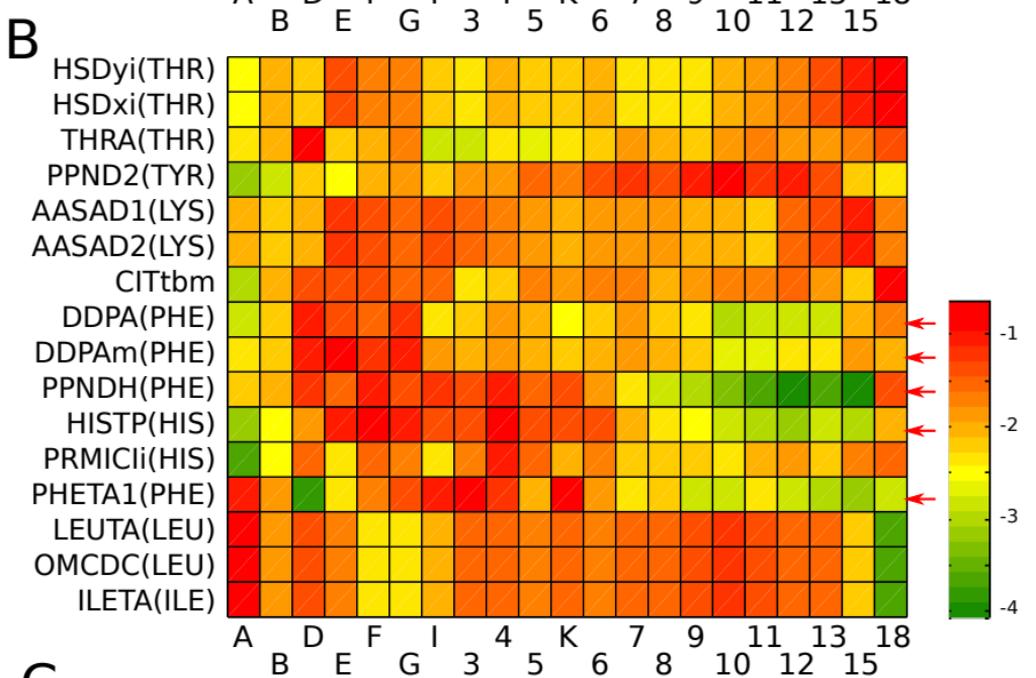
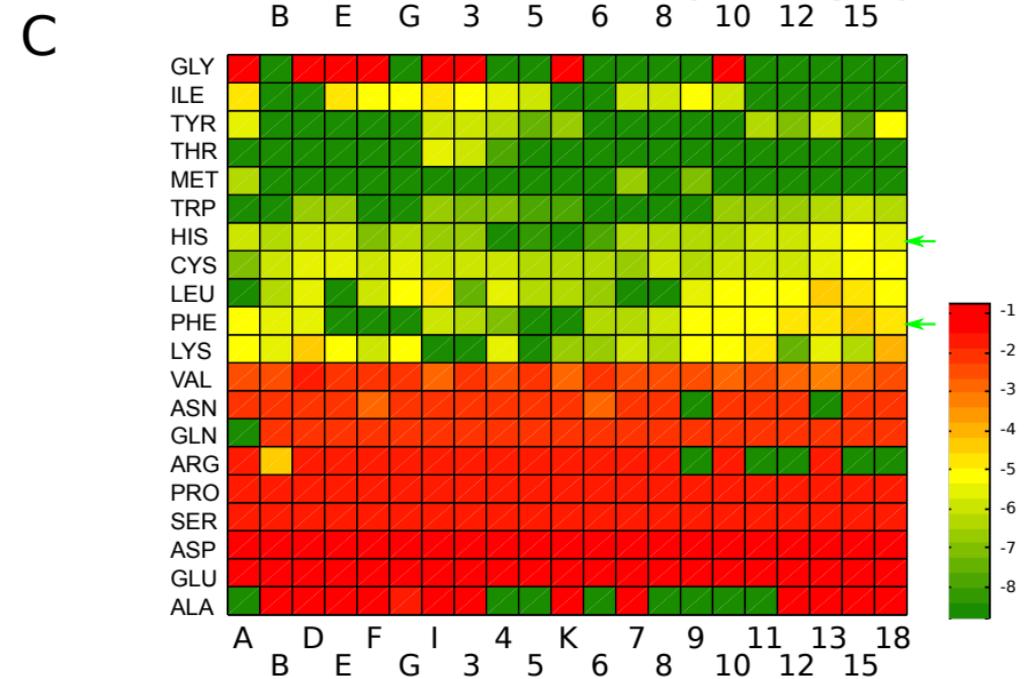